\documentclass[sigconf]{acmart}

\usepackage{textcomp}
\usepackage{stfloats}
\usepackage{url}
\usepackage{verbatim}
\usepackage{graphicx}

\usepackage{tikz}
\usepackage{comment}
\usepackage{color}

\usepackage{enumitem}
\usepackage{graphicx}
\usepackage{amsmath}
\usepackage{booktabs}
\usepackage{mathtools}
\usepackage{url}
\usepackage{comment}
\usepackage{color}
\usepackage{graphicx,setspace}
\usepackage{multirow, array, booktabs,makecell,xspace,bm}
\usepackage{pifont}
\usepackage{amsfonts}
\usepackage{arydshln}
\usepackage{dsfont}

\usepackage{stackengine}

\usepackage{scalerel}
\usepackage{esdiff} 

\usepackage{bbm}

\newcommand{\cmark}{\ding{51}}%

\AtBeginDocument{%
	}

\copyrightyear{2023}
\acmYear{2023}
\setcopyright{acmlicensed}\acmConference[MM '23]{Proceedings of the 31st ACM International Conference on Multimedia}{October 29-November 3, 2023}{Ottawa, ON, Canada}
\acmBooktitle{Proceedings of the 31st ACM International Conference on Multimedia (MM '23), October 29-November 3, 2023, Ottawa, ON, Canada}
\acmPrice{15.00}
\acmDOI{10.1145/3581783.3611722}
\acmISBN{979-8-4007-0108-5/23/10}

\begin{document}
	
	\title{Audio-Visual Spatial Integration and Recursive Attention for Robust Sound Source Localization}
	
	\author{Sung Jin Um}
	\authornote{Both authors have contributed equally to this work.}
	\email{sungzin1@khu.ac.kr}
	\affiliation{%
		\institution{Kyung Hee University}
		\city{Yongin-si}
		\country{South Korea}
	}
	
	\author{Dongjin Kim}
	\email{rlaehdwls310@khu.ac.kr}
	\authornotemark[1]
	\affiliation{%
		\institution{Kyung Hee University}
		\city{Yongin-si}
		\country{South Korea}
	}
	
	\author{Jung Uk Kim}
	\authornote{Corresponding author.}
	\email{ju.kim@khu.ac.kr}
	\affiliation{%
		\institution{Kyung Hee University}
		\city{Yongin-si}
		\country{South Korea}
	}

	\begin{abstract}
		The objective of the sound source localization task is to enable machines to detect the location of sound-making objects within a visual scene. While the audio modality provides spatial cues to locate the sound source, existing approaches only use audio as an auxiliary role to compare spatial regions of the visual modality. Humans, on the other hand, utilize both audio and visual modalities as spatial cues to locate sound sources. In this paper, we propose an audio-visual spatial integration network that integrates spatial cues from both modalities to mimic human behavior when detecting sound-making objects. Additionally, we introduce a recursive attention network to mimic human behavior of iterative focusing on objects, resulting in more accurate attention regions. To effectively encode spatial information from both modalities, we propose audio-visual pair matching loss and spatial region alignment loss. By utilizing the spatial cues of audio-visual modalities and recursively focusing objects, our method can perform more robust sound source localization. Comprehensive experimental results on the Flickr SoundNet and VGG-Sound Source datasets demonstrate the superiority of our proposed method over existing approaches. Our code is available at: \href{https://github.com/VisualAIKHU/SIRA-SSL}{\color{red}{https://github.com/VisualAIKHU/SIRA-SSL}}.
	\end{abstract}

	\ccsdesc[500]{Information systems~Multimedia information systems}
	\ccsdesc[500]{Computing methodologies~Computer vision}

	\keywords{Sound source localization, audio-visual spatial integration, recursive attention, multimodal learning}

	\maketitle

	\section{Introduction}
	
	Sound source localization aims to identify the location of a sounding object within a visual scene \cite{CVPR2018_Senocak}. This task is similar to the innate ability of humans to find the location by correlating sounds heard with their ears and scenes seen with their eyes. Because of this property, sound source localization has a wide range of applications, such as multimodal robotics \cite{IEEE2016_Li, kim2023towards}, sound source separation \cite{IEEE2019_chazan}, and indoor positioning \cite{ASA2017_Gar}. 
	
	\begin{figure}[t]
		\begin{minipage}[b]{1.0\linewidth}
			\centering
			\centerline{\includegraphics[width=8.8cm]{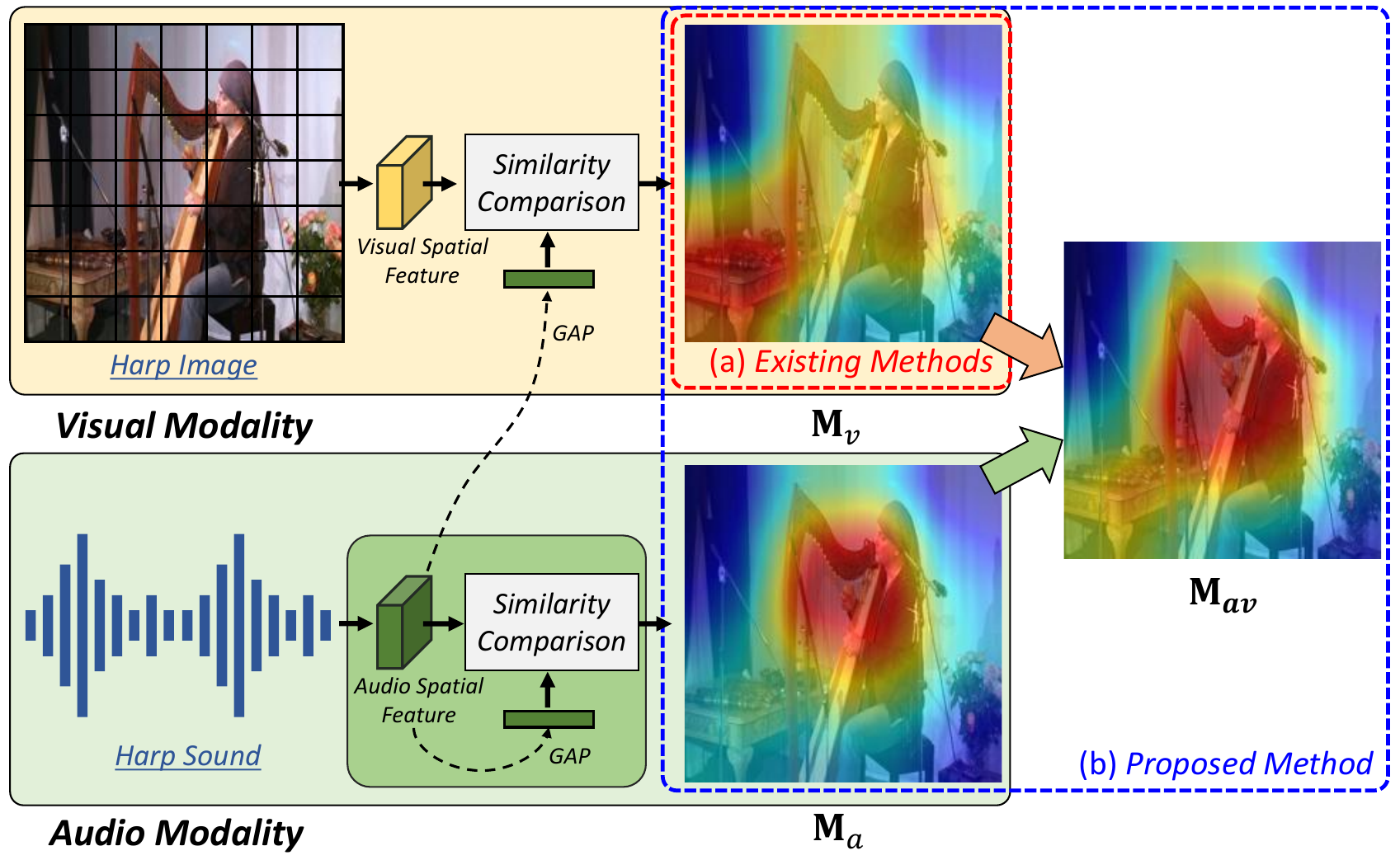}}
		\end{minipage}
		\vspace{-0.4cm}
		\caption{Conceptual comparison between (a) existing methods (red) and (b) the proposed method (blue). The existing methods use the spatial information of visual modality as the primary modality to estimate region of sound-making objects ($\textbf{M}_{v}$). We observe that the audio modality itself also contains spatial information for estimating regions of the sound-making object ($\textbf{M}_{a}$). In our work, we try to integrate the spatial knowledge of the audio-visual modalities ($\textbf{M}_{av}$) for more accurate sound source localization.}
		\vspace{-0.2cm}
		\label{fig:1}
	\end{figure}
	
	Since the sound source localization task utilizes multimodal information (\textit{i.e.,} audio-visual), it is essential to consider how to effectively combine the different two modal information for more accurate localization. In addition, while audio-visual data can be obtained in abundance, manually annotating object locations (\textit{e.g.,} bounding boxes or segmentation masks) is time-consuming and labor-intensive. To address the two issues, several self-supervised approaches \cite{CVPR2018_Senocak,CVPR2021_Chen,CVPR2022_Xuan,WACV2023_Fedorishin,WACV2022_Shi,WACV2023_Zhou} have been proposed. Senocak et al. \cite{CVPR2018_Senocak} proposed the attention mechanism with unsupervised learning to match the audio-visual information. Chen et al. \cite{CVPR2021_Chen} introduced a network to explicitly mine the hard negative locations from the foreground locations by using sound information. Xuan et al. \cite{CVPR2022_Xuan} proposed a proposal-based method that focuses on the region inside the bounding box of each object based on the given sound. In \cite{WACV2023_Fedorishin}, the optical flow information was additionally incorporated to effectively combine the audio-visual modalities.
	
	However, the above-mentioned methods have in common that, as shown in Figure \ref{fig:1}(a), they utilize the audio modality only as an auxiliary role (red) in comparing whether each grid region of the visual modality corresponds to the area of the sounding object. In fact, humans also have the ability to detect the location of an object just by hearing the sound. For example, even when our eyes are closed, we can still perceive the location of a car making a sound by paying our attention to the corresponding spatial area. This is because the spatial information can be inferred by relying on cues such as differences in arrival time, loudness, and spectral content of the sound \cite{shelton1980influence, gaver1993world, majdak2010_sound}. As shown in Figure \ref{fig:1}(b), we observe that the audio modal itself also contains valuable spatial cues for inferring the sound-making objects.
	
	Moreover, according to \cite{jones1975eye, perrott1996aurally, bolia1999aurally}, when humans receive both visual and auditory information, they naturally generate a region of interest (ROI) in each modality. These ROIs are then integrated to form a region of attention, which is an indicator of where to focus based on the combined audio-visual information. After focusing the attention region and eliminating the unnecessary areas, humans identify the sound-making object by repeatedly engaging in a recursive recognition process \cite{luo1989multisensor,itti2001computational}. By doing so, we can make more accurate predictions. This cognitive process enables humans to effectively utilize visual and auditory information, leading to more accurate and comprehensive understanding of the world around them.
	
	In this paper, based on our aforementioned motivations, we propose a novel sound source localization framework that mimics the above-mentioned two cognitive psychological perspectives of humans (\textit{i.e.,} potential of spatial cues in audio modality and the ability to recognize sound-making objects in a recursive manner). Our framework consists of two stages. First, we propose an audio-visual spatial integration network that integrates spatial knowledge from both audio-visual modalities to produce an integrated localization map. The aim of generating the integrated localization map is to contain rich spatial information about the sound-making objects. Second, we introduce a recursive attention network to mimic the human ability to recognize the objects in a recursive manner. Based on the integrated localization map, the unnecessary regions of the input image are eliminated and attentive input image is generated. Consequently, with the attentive input image, more precise localization of the sound-making object is possible in our recursive attention network.
	In addition, within the recursive attention network, we devise an audio-visual pair matching loss to guide the feature representation of each single modality (audio and visual) to resemble that of the attentive input image. By doing so, the features of both modalities can embed more precise spatial knowledge. Moreover, although the spatial knowledge of the audio modality contains valuable information, it may be relatively less precise than those of the visual modality. To address this issue, we introduce a spatial region alignment loss to guide the spatial representation of the audio modality to resemble that of the attentive input image. As a result, the feature representations of the audio modality are significantly enhanced, leading to a more accurate final localization map generation.
	
	To sum up, the major contributions of this paper are summarized as follows:
	\vspace{0.1cm}
	\begin{itemize}
		\item We introduce audio-visual spatial integration network that exploits the spatial knowledge of audio-visual modalities. In addition, we propose recursive attention network to refine the localization map in a recursive manner. To the best of our knowledge, it is the first work that considers the spatial knowledge of audio modality for sound source localization.
		\vspace{0.45cm}
		\item To guide the feature representation of the single modality, we propose audio-visual pair matching loss. Also, to enhance spatial knowledge of the audio modality, we introduce spatial region alignment loss to resemble that of the attentive image.
		\vspace{0.01cm}
		\item Comprehensive quantitative and qualitative experimental results on Flickr-SoundNet and VGG-Sound Source datasets validate the effectiveness of the proposed framework.
	\end{itemize}
	
	\section{Related Work}
	
	\subsection{Sound Source Localization}
	
	Sound source localization aims to estimate the sound source location using visual scenes. It requires an effective combination of visual and audio data, and various algorithms have been developed over the years to optimize this multimodal integration for accurate localization \cite{CVPR2018_Senocak, CVPR2021_Chen, CVPR2022_Xuan, WACV2023_Fedorishin, CVPR2019_Hu_dmc, WACV2023_Zhou, WACV2022_Shi, ECCV2020_afouras_avobject, ECCV2020_Qian_coarsetofine}. 
	
	One such approach is the use of attention mechanisms, which allow the network to selectively focus on relevant parts of the input data. In \cite{CVPR2018_Senocak}, Senocak et al. propose a sound localization network that incorporates an attention mechanism to focus on relevant parts of the visual modality and audio modality, resulting in more accurate sound source localization. In \cite{CVPR2021_Chen}, Chen et al. introduce tri-maps to incorporate background mining techniques for identifying positive correlation region, no correlation region (background), and ignoring region to avoid uncertain areas in the visual scene. They utilize audio-visual pairs to create a tri-map highlighting positive/negative regions. In \cite{CVPR2022_Xuan}, Xuan et al. adopt the selective search \cite{uijlings2013selective} to utilize the proposal-based paradigm. Since the proposal region contains information of sound-making objects, finding the candidate objects firstly rather than the location of the sound can be superior. In \cite{WACV2023_Fedorishin}, Fedorishin et al., assumed that most of the sound sources in visual scenes will be moving objects. Therefore, they adopt the optical flow algorithm in the visual modality to achieve more effective sound source localization.
	
	In many studies on sound source localization task, the visual modality is usually considered to be a crucial modality (\textit{e.g.,} selective search, optical flow, etc.). However, the audio modality is only utilized as an auxiliary role, primarily being used for similarity measurements  (\textit{e.g.,} cosine similarity) to generate the attention region of the visual modality. Thus, we claim that the existing methods tend to give weight to visual modality rather than audio modality. However, humans use both eyes and ears as important factors to judge situations in the natural environment. Therefore, we propose a sound source localization framework that uses audio modality as well as visual modality for acquire more abundant spatial knowledge of the audio-visual modalities.
	
	\begin{figure*}[t]
		\begin{minipage}[b]{1.0\linewidth}
			\centering
			\centerline{\includegraphics[width=17.9cm]{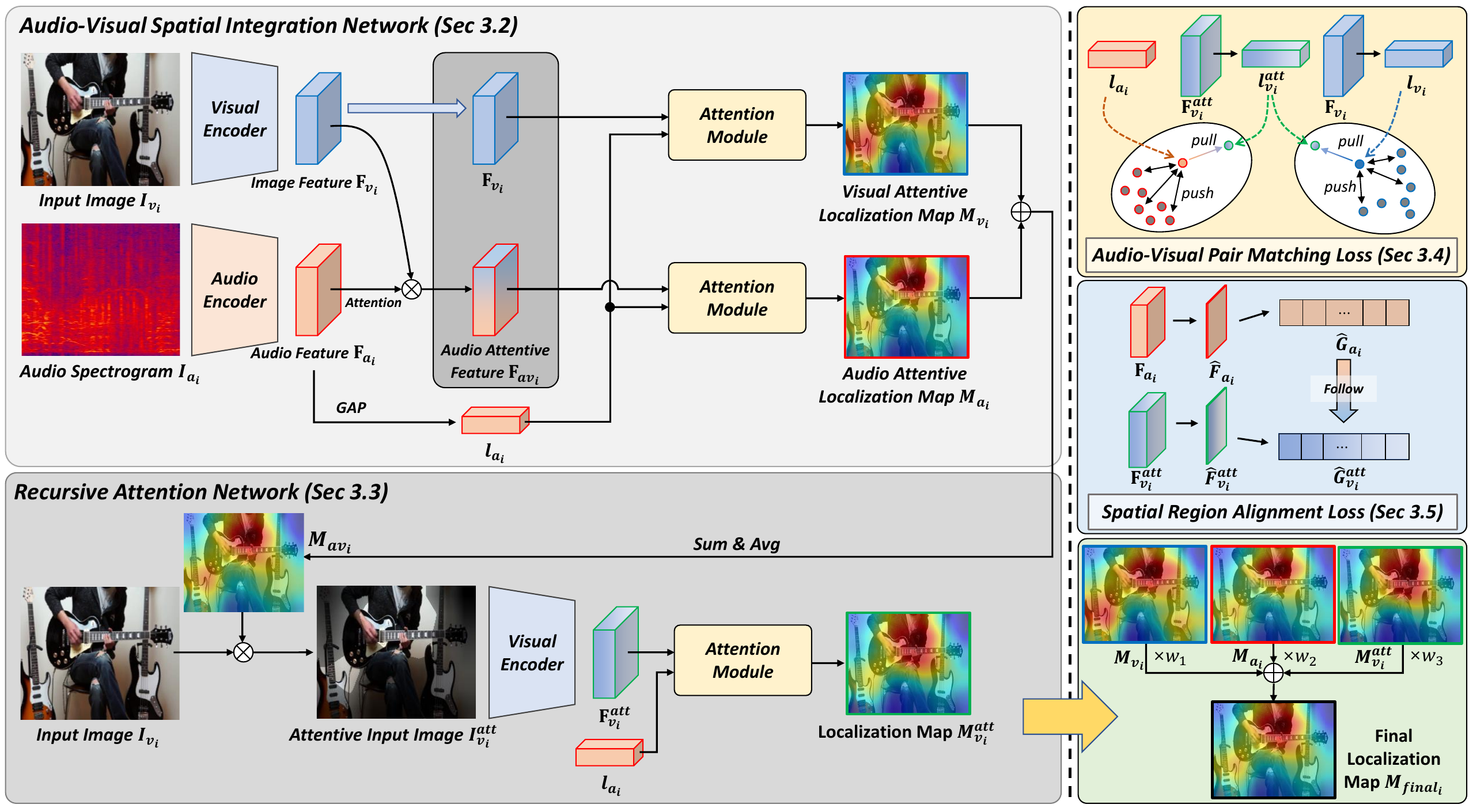}}
		\end{minipage}
		\vspace{-0.8cm}
		\caption{Network configuration of the proposed sound localization framework. $\oplus$ and $\otimes$ indicate element-wise addition and element-wise multiplication, respectively. Note that final localization map $\textbf{M}_{final}$ is generated by combining $\textbf{M}_{v}$, $\textbf{M}_{a}$, and $\textbf{M}_{v}^{att}$.}
		\vspace{-0.2cm}
	\end{figure*}
	
	\subsection{Recursive Deep Learning Framework in Computer Vision}
	
	Recursive deep learning frameworks \cite{al2017aroma, kim2017iterative, li2014recursive, socher2014recursive, irsoy2014deep, socher2011parsing} have become increasingly popular for their ability to handle complex dependencies in sequential or structured data. Many works have adopted a recursive approach and applied it to the various computer vision tasks to improve their performance, such as object detection \cite{li2017multistage_recursive, ke2022recursive_object_detection, chen2021recursive_object_detection} and recognition \cite{socher2012recursive_object_recognition, bui2016object_recognition, bui2016using_object_recognition}, image super-resolution \cite{kim2016deeply_resolution, wei2020deep_resolution, tai2017image_resolution}, visual tracking \cite{gao2020recursive_visualtracking, huang2019bidirectional_visualtracking}, and semantic segmentation \cite{zhang2018recursive_segmentation, zhang2020deep_segmentation, pinheiro2014recurrent_segmentation}.
	
	For example, in the object detection, a recursive model with the multistage framework is proposed \cite{li2017multistage_recursive}. This approach uses an EM-like group recursive learning technique to iteratively refine object proposals and improve the spatial configuration of object detection. Socher et al. \cite{socher2012recursive_object_recognition} proposed a model that combines convolutional and recursive neural networks to detect object in the RGB-D images. In addition, for image super-resolution, Kim et al. \cite{kim2016deeply_resolution} proposed the deeply-recursive convolutional network (DRCN) to improve the feature representation without adding more convolution parameters. To overcome the challenges of learning a DRCN, they introduce recursive supervision and skip connection.
	
	In the visual object tracking, Gao et al. \cite{gao2020recursive_visualtracking} utilized recursive least-squares estimation (LSE) for online learning. By integrating fully-connected layers with LSE and employing an enhanced mini-batch stochastic gradient descent algorithm, they enhanced the performance of visual object tracking. For semantic segmentation and depth estimation tasks, Zhang et al. \cite{zhang2018recursive_segmentation} introduced the Joint Task-Recursive Learning (TRL) framework. It uses a Task-Attentional Module (TAM) to recursively refine the results.
	
	For designing our method, we utilize the recursively refining idea to mimic the behavior of humans that repeatedly focus sound-making object for more accurate sound source localization. By recursively refining a model, the proposed method can improve the attention region of the sound-making object, by eliminating the unnecessary regions. As a result, our method achieves the outstanding performance over the state-of-the-art sound source localization works.
	
	\section{Proposed Method}
	
	\subsection{Overall Architecture}
	
	The overall architecture of our sound source localization framework is depicted in Figure 2. Our framework consists of two stages: (1) audio-visual spatial integration network and (2) recursive attention network. First, in the audio-visual spatial integration network, input image set $I_v\in\mathbb{R}^{N\times W_v \times H_v \times 3}$ ($N$ indicates the number of batch, $W_v$ and $H_v$ denote width and height of $I_v$, respectively) and the corresponding audio spectrogram set $I_a\in\mathbb{R}^{N\times W_a\times H_a \times 1}$ ($W_a$ and $H_a$ denote width and height of $I_a$, respectively) pass through each modal encoder (\textit{i.e.,} visual encoder and audio encoder) to generate the spatial features $\textbf{F}_v$ and $\textbf{F}_a$, respectively. Then, image attentive localization map $\textbf{M}_v$ and audio attentive localization map $\textbf{M}_a$ are generated based on $\textbf{F}_v$ and $\textbf{F}_a$ through the attention module. $\textbf{M}_v$ and $\textbf{M}_a$ are attention maps that focus on the location of a sounding object based on the spatial features encoded in each modality. $\textbf{M}_v$ and $\textbf{M}_a$ are integrated to generate the audio-visual integrated localization map $\textbf{M}_{av}$.
	
	Second, the recursive attention network takes the resized $\textbf{M}_{av}$ and multiplies it with $I_v$ to generate attentive input image $I_v^{att}$. $I_v^{att}$ is passed through the visual encoder to generate visual attention feature $\textbf{F}_v^{att}$. Note that the weight parameters of the visual encoder in the audio-visual spatial integration network and recursive attention network are shared. With $\textbf{F}_v^{att}$ and $I_a$, the localization map $\textbf{M}_v^{att}$ is generated. More details are in the following subsections.
	
	\subsection{Audio-Visual Spatial Integration Network}
	
	When humans see a visual scene with their eyes while listening to a sounding object, they can acquire spatial cue information not only through vision but also through sound \cite{shelton1980influence, gaver1993world}. We mimic the behaviors of humans for more accurate localizing sound source objects. To this end, we propose an audio-visual spatial integration network to exploit the spatial cues of both visual modality and audio modalities.

	As shown in Figure 2, our audio-visual spatial integration network consists of two streams: (1) visual stream and (2) audio stream. In the visual stream, the visual spatial feature $\textbf{F}_v\in\mathbb{R}^{N\times w\times h\times c}$ ($w$, $h$, and $c$ are the width, height, and channel number) is mainly used to localize sound-making object. Specifically, the audio spatial feature $\textbf{F}_a\in\mathbb{R}^{N\times w\times h \times c}$ is subject to a global average pooling (GAP) operation to generate $l_a\in\mathbb{R}^{N\times c}$. Then, $\textbf{F}_v$ and $l_a$ are compared using a similarity calculation in the attention module to generate $\textbf{S}_v=\{S_{v_{ij}}\}_{i=1,...,h, j=1,...,w}\in\mathbb{R}^{N\times w \times h}$, which is measured as:
	\begin{equation}
		\begin{gathered}
			S_{v_{ij}} = \frac{Sim({\textbf{F}}_{v_{ij}},l_a)}{\sum_{i=1}^{h}\sum_{j=1}^{w}Sim({\textbf{F}}_{v_{ij}},l_a)}, \,\, Sim({\textbf{F}}_{v_{ij}},l_a)=\frac{{\textbf{F}}_{v_{ij}} \cdot l_a} {{||{\textbf{F}}_{v_{ij}}||}\,{||l_a||}}.
		\end{gathered}
	\end{equation}
	Then, $\textbf{S}_v$ is normalized by the softmax to generate the image attentive localization map $\textbf{M}_v\in\mathbb{R}^{N\times w \times h}$.
	
	In the audio stream, the audio spatial feature $\textbf{F}_a\in\mathbb{R}^{N\times w\times h \times c}$ is mainly used to localize sound-making objects. However, while the audio modality contains the spatial cues for localizing objects, it generally lacks the levels of detail compared to the visual modality. For example, if we hear an object sound with our eyes closed, we can roughly estimate its location, but it is typically less precise than if we were to open our eyes and visually locate the object. Thus, we transfer the spatial knowledge of $\textbf{F}_v$ to $\textbf{F}_a$ while maintaining the area that $\textbf{F}_a$ focuses on by generating $\textbf{F}_{av}$. $\textbf{F}_{av}$ is obtained as:
	\begin{equation}
		\textbf{F}_{av} = \textbf{F}_v \circ \bar{\textbf{F}}_a,
	\end{equation}
	where $\bar{\textbf{F}}_a$ denotes the normalized version of $\textbf{F}_a$ (min-max normalization is conducted with the value between 0 and 1), and $\circ$ indicates the element-wise multiplication.
	
	Next, similar to Eq. (1), the  $\textbf{S}_{av}=\{S_{{av}_{ij}}\}_{i=1,...,h, j=1,...,w}\in\mathbb{R}^{N\times w \times h}$ is obtained as:
	\begin{equation}
		\begin{gathered}
			S_{{av}_{ij}} = \frac{Sim({\textbf{F}}_{{av}_{ij}},l_a)}{\sum_{i=1}^{h}\sum_{j=1}^{w}Sim({\textbf{F}}_{{av}_{ij}},l_a)}, \,\, Sim({\textbf{F}}_{{av}_{ij}},l_a)=\frac{{\textbf{F}}_{{av}_{ij}} \cdot l_a} {{||{\textbf{F}}_{{av}_{ij}}||}\,{||l_a||}}.
		\end{gathered}
	\end{equation}
	$\textbf{S}_{av}$ is also normalized by the softmax to make the audio attentive localization map $\textbf{M}_{a}$. The two localization maps, $\textbf{M}_v$ and $\textbf{M}_a$, generated by the proposed audio-visual spatial integration network, provide information about the spatial regions in each modality that are being focused on to localize the sounding objects. Therefore, we integrate the knowledge of the audio-visual modalities to make $\textbf{M}_{av}\in\mathbb{R}^{N\times w \times h}$, which can be obtained as follows:
	\begin{equation}
		\textbf{M}_{av}=\frac{\textbf{M}_{a}+\textbf{M}_{v}}{2}.
	\end{equation}
	Since $\textbf{M}_{av}$ contains the spatial information of both audio and visual modalities, it provides a more precise localization map compared to using either modality alone. By combining the spatial cues from both modalities, the proposed method is able to effectively mitigate the limitations of each modality and produce a more accurate localization result.
	
	\subsection{Recursive Attention Network}
	
	Given the visual and audio modal information, humans can integrate attention regions across different modalities, such as visual and auditory information, to concentrate on a specific region \cite{MultisensoryIntegration,driver1998cross,eimer1998erp,xu2022ava}. It is called multisensory integration. By doing so, humans can concentrate their attention on specific regions of the environment that correspond to the presented sensory information. This allows them to more effectively process and respond to stimuli from both modalities \cite{mesulam1981cortical,foxe2000multisensory,lee2022audio}.
	
	Therefore, we build the recursive attention network to mimic the above-mentioned behaviors of humans. The recursive attention network utilizes the audio-visual integrated localization map $\textbf{M}_{av}$ derived from the audio-visual spatial integration network to produce an attentive input image $I_v^{att}$. Specifically, $\textbf{M}_{av}\in\mathbb{R}^{N\times w \times h}$ is resized to be $\textbf{M}_{av}^r\in\mathbb{R}^{N\times W_v \times H_v}$. To the next, $\textbf{M}_{av}^r$ and $I_v$ are conducted element-wise multiplication to focus the attention region of the image, \textit{i.e.,}  $I_v^{att}$. We feed this attentive input image $I_v^{att}$ into the visual encoder to encode visual attention feature $F_v^{att}$. The attention module calculates the similarity between $F_v^{att}$ and $l_a$ to generate $\textbf{S}_v^{att}=\{S_{v_{ij}}^{att}\}_{i=1,...,h, j=1,...,w}\in\mathbb{R}^{N\times w \times h}$. Note that $\textbf{S}_v^{att}$ is calculated similarly to Eq. (1) and Eq. (3). Also, $\textbf{S}_v^{att}$ is normalized by the softmax to make the localization map $\textbf{M}_v^{att}$.

	Finally, we combine the $\textbf{M}_v$, $\textbf{M}_a$, and $\textbf{M}_v^{att}$ to generate the final localization map $\textbf{M}_{final}$, which can be represented as:
	\begin{equation}
		\textbf{M}_{final} = {w_1}\textbf{M}_{v} + {w_2}\textbf{M}_a + {w_3}\textbf{M}_v^{att},
	\end{equation}
	where $w_1$, $w_2$, and $w_3$ are the hyper-parameters that indicate the importance of each modality in contributing to the $\textbf{M}_{final}$. $\textbf{M}_{a}$ and $\textbf{M}_{v}$ contain the spatial cues of each modality (\textit{i.e.,} audio and visual modalities), and $\textbf{M}_v^{att}$ contains the spatial cues of the more attentive region from the audio-visual modalities. Therefore, by combining $\textbf{M}_{a}$ and $\textbf{M}_{v}$, the spatial cues from both modalities can be obtained. Additionally, by combining $\textbf{M}_v^{att}$, more interested regions can be obtained. The recursive combination of the localization maps can utilize abundant spatial information, leading to more accurate sound source localization.
	
	\subsection{Audio-Visual Pair Matching Loss}
	
	Humans can make more accurate predictions by removing unnecessary areas by focusing attention through their eyes and ears. Similarly, in our method, the attentive input image $I_v^{att}$ concentrates the area that is generated by the audio-visual modality in the audio-visual spatial integration network. This enables us to localize the sounding objects more accurately. This is similar to the fact that the two-stage detectors \cite{ren2015faster,kim2021robust,ghiasi2019fpn,lin2017feature,kim2021uncertainty}, which first extract the region of interest (ROI) for more accurate object detection, generally outperform the one-stage object detectors \cite{redmon2016you,liu2016ssd,lin2017focal}. Therefore, compared to $\textbf{M}_v$, $\textbf{M}_a$ and $\textbf{M}_v^{att}$, $\textbf{M}_v^{att}$ usually contains more meaningful regions than $\textbf{M}_v$ and $\textbf{M}_a$. As a result, we propose an audio-visual pair matching loss to guide the feature representations of the visual modality $\textbf{F}_v$ and the audio modality $\textbf{F}_a$ to be similar to that of the visual attention feature $\textbf{F}_v^{att}$.
	
	To this end, we first conduct global average pooling (GAP) of $\textbf{F}_v^{att}$, $\textbf{F}_v$, and $\textbf{F}_a$ and normalize them to generate $l_{v}^{att}$, $l_{v}$, and $l_a$, respectively. Next, we adopt the triplet loss \cite{triplet-loss_2015} for the audio-visual pair matching loss $\mathcal{L}_{avpm}$, which can be represented as:
	\begin{equation}
		\begin{gathered}
			T(l_{v_i}^{att},l_{a_i},l_{a_j}) = D(l_{v_i}^{att}, l_{a_i}) + max(\delta - D(l_{v_i}^{att},l_{a_j}), 0), \\
			T(l_{v_i}^{att},l_{v_i},l_{v_j}) = D(l_{v_i}^{att}, l_{v_i}) + max(\delta - D(l_{v_i}^{att},l_{v_j}), 0),\\
			\mathcal{L}_{avpm}=\frac{1}{N(N-1)}\sum^N_{i=1}\sum^N_{j=1(j\neq i)}T(l_{v_i}^{att},l_{a_i},l_{a_j}) + T(l_{v_i}^{att},l_{v_i},l_{v_j}),
		\end{gathered}
	\end{equation}
	where $D(\alpha,\beta) = ||(\alpha - \beta)/\tau||_2^2$ denotes the L2 norm to calculate the distance between two features with temperature parameter $\tau$, $l_{v_i}^{att}$, $l_{a_i}$, and $l_{a_j}$ are the features of anchor, positive, and negative samples, respectively, and $\delta$ is the margin.
	
	The aim of $T(l_{v_i}^{att},l_{a_i},l_{a_j})$ and $T(l_{v_i}^{att},l_{v_i},l_{v_j})$ is to make the anchor ($l_{v_i}^{att}$) and the positive pair ($l_{a_i}$, $l_{v_i}$) similar while pushing the negative pair ($l_{a_j}$, $l_{v_j}$) apart. By doing so, $\mathcal{L}_{avpm}$ can guide the feature representation of $\textbf{F}_v$ and $\textbf{F}_a$ to be similar that of $\textbf{F}_v^{att}$. As a result, the feature representation of $\textbf{F}_v$ and $\textbf{F}_a$ improve the performance of sound source localization (please see Section 4.5).
	
	\subsection{Spatial Region Alignment Loss}
	
	Although we can infer spatial information using sound, it is relatively less accurate than visual information. Therefore, we introduce a spatial region alignment loss in order to guide the spatial regions that audio feature $\textbf{F}_a$ focus on to be similar to that of the $\textbf{F}_v^{att}$. To this end, we add all $c$ channels of $\textbf{F}_a$ and $\textbf{F}_v^{att}$ to normalize them to generate $\hat{\textbf{F}}_a\in\mathbb{R}^{N\times w\times h}$ and $\hat{\textbf{F}}_v^{att}\in\mathbb{R}^{N\times w\times h}$. After that, they are flattened to conduct softmax function to generate $\hat{\textbf{G}}_a\in\mathbb{R}^{N\times wh}$ and $\hat{\textbf{G}}_v^{att}\in\mathbb{R}^{N\times wh}$, respectively. Based on the $\hat{\textbf{G}}_v^{att}$ and $\hat{\textbf{G}}_a$, the spatial region alignment loss $\mathcal{L}_{sra}$ is represented as follows:
	\begin{equation}
		\begin{aligned}
			\mathcal{L}_{sra}=
			\frac{1}{N} \sum_{i=1}^{N} \underbrace{D_{KL}\left(\hat{\textbf{G}}_{v_i}^{att} \Vert \hat{\textbf{G}}_{a_i}\right)}_\text{audio to attentive visual},
		\end{aligned}
	\end{equation}
	where $D_{KL}(\cdot)$ indicates the Kullback-Leibler (KL) divergence. $\mathcal{L}_{sra}$ makes the spatial representation of $\textbf{F}_a$ to be similar to that of $\textbf{F}_v^{att}$ in the training phase. By doing so, when generating $\textbf{F}_a$, our method can effectively estimate the spatial regions by hearing sounds.
	
	\subsection{Total Loss Function}
	To train our method, the total loss function is composed as follows: 
	\begin{equation}
		\mathcal{L}_{Total}=\mathcal{L}_{SSL}+\lambda_1\mathcal{L}_{avpm}+\lambda_2\mathcal{L}_{sra},
	\end{equation}
	where $\mathcal{L}_{SSL}$ is the unsupervised loss function of the sound source localization that tries to impose the audio-visual feature pairs are close to each other, following \cite{CVPR2021_Chen}, $\lambda_1$ and $\lambda_2$ denote the balancing parameter. Through $\mathcal{L}_{Total}$, our method can perform effective sound source localization by leveraging the spatial knowledge of the audio-visual modality and combining all attention maps in a recursive manner.
	
	\begin{table}[t]
		\renewcommand{\tabcolsep}{0.6mm}
		\centering
		\caption{Experimental results on Flickr test set when the training sets are Flickr10k and Flickr144k, respectively.}
		\vspace{-0.3cm}
		\resizebox{0.9999\linewidth}{!}{
			\begin{tabular}{cccc}
				\Xhline{3\arrayrulewidth}
				\rule{0pt}{9.5pt} \bf Method & \bf Training Set & \bf cIoU$_{0.5}$$\uparrow$ & \bf AUC$\uparrow$ \\ \hline
				\rule{0pt}{9pt}
				Attention \cite{CVPR2018_Senocak} (CVPR'18) & \multirow{9}{*}{\bf Flickr10k}  & 0.436 & 0.449  \\
				DMC \cite{CVPR2019_Hu_dmc} (CVPR'19)        &  & 0.414 & 0.450  \\
				CoarseToFine \cite{ECCV2020_Qian_coarsetofine} (ECCV'20)      &  & 0.522 & 0.496  \\
				AVObject \cite{ECCV2020_afouras_avobject} (ECCV'20)          &  & 0.546 & 0.504  \\
				LVS \cite{CVPR2021_Chen} (CVPR'21)       &  & 0.582 & 0.525  \\
				Zhou et al. \cite{WACV2023_Zhou} (WACV'23)       &  & 0.631 & 0.551  \\
				Shi et al. \cite{WACV2022_Shi} (WACV'22) &  & 0.734 & 0.576  \\
				SSPL \cite{CVPR2022_Xuan} (CVPR'22)      &  & 0.743 & 0.587  \\
				HTF \cite{WACV2023_Fedorishin} (WACV'23)           &  & 0.860 & 0.634  \\\cdashline{1-4}
				\rule{0pt}{9.5pt}
				\bf Proposed Method        &  & \bf 0.876 & \bf 0.641 \\
				\Xhline{2\arrayrulewidth}
				\rule{0pt}{9.pt}
				Attention \cite{CVPR2018_Senocak} (CVPR'18) & \multirow{5}{*}{\bf Flickr144k}  & 0.660 & 0.558  \\
				DMC \cite{CVPR2019_Hu_dmc} (CVPR'19)        &  & 0.671 & 0.568  \\
				LVS \cite{CVPR2021_Chen} (CVPR'21)       &  & 0.699 & 0.573  \\
				SSPL \cite{CVPR2022_Xuan} (CVPR'22)      &  & 0.759 & 0.610  \\
				HTF \cite{WACV2023_Fedorishin} (WACV'23)       &  & 0.865 & 0.639  \\\cdashline{1-4}
				\rule{0pt}{9.5pt}
				\bf Proposed Method       &  & \bf 0.881 & \bf 0.652 \\
				\Xhline{3\arrayrulewidth}
			\end{tabular}
		}
		\label{table:param}
		\vspace{-0.6cm}
	\end{table}
	
	\section{Experiments}
	
	\subsection{Datasets and Evaluation Metrics}
	
	\noindent {\textbf{Flickr-SoundNet.}} Flickr-SoundNet \cite{flickr} consists more than 2 million videos from Flickr. In the training phase, to enable direct comparison with prior research, we train our models with two random subsets of 10k and 144k image-audio pairs. In the inference phase, we use Flickr-SoundNet test set. It contains 250 annotated pairs with labeled bounding box, manually annotated by the annotators \cite{CVPR2018_Senocak,CVPR2021_Chen}. \\
	
	\noindent {\textbf{VGG-Sound Source.}} VGG-Sound dataset \cite{chen2020vggsound} consists of 200k video clips from 300 different sound categories. Following \cite{WACV2023_Fedorishin}, we use a training dataset with 10k and 144k image-audio pairs. For evaluation, we use VGG-Sound Source (VGG-SS) dataset \cite{CVPR2021_Chen} with 5,000 annotated image-audio pairs from 220 classes. Compared with Flickr-SoundNet, which contains 50 audio categories, the VGG-SS dataset set offers a larger number of sound sources. Therefore, it contains more challenging scenario for sound source localization. \\
	
	\noindent {\textbf{Evaluation Metrics.}} To compare our method with the existing methods, we adopt consensus Intersection over Union (cIoU) \cite{CVPR2018_Senocak} and Area Under Curve (AUC) as evaluation metrics, which are the widely adopted metrics for sound source localization task \cite{CVPR2018_Senocak,CVPR2021_Chen,WACV2023_Fedorishin}. For calculating cIoU, the IoU threshold is fixed to be 0.5 (\textit{i.e.,} cIoU$_{0.5}$), following \cite{CVPR2018_Senocak, CVPR2021_Chen, WACV2023_Fedorishin}. Note that, in our experiments, we additionally introduce a mcIoU metric to measure the performance by varying the IoU threshold to 0.5:0.05:0.95. More details are in Section 4.6. 
	
	\subsection{Implementation Details}
	
	For both datasets, we resize the input image for the visual modality to be $W_v=224$, $H_v=224$. It is extracted from the middle frame of the 3-seconds video clips. For audio modality input, we resample the 3-seconds raw audio signal to 16kHz and transform it into a log-scale spectrogram, yielding a final shape $W_a=257$ and $H_a=276$. At this time, to enable a direct comparison with visual modal features, we resize $\textbf{F}_a$ to be $7\times 7\times 512$ ($w=7$, $h=7$, and $c=512$) using bilinear interpolation.
	
	Following \cite{CVPR2021_Chen}, we employ ResNet-18 \cite{resnet} for both the visual and audio feature backbones to construct our baseline. Since the number of sound spectrogram channel is 1. we modify the input channel of ResNet-18 \cite{resnet} conv1 from 3 to 1. We use the ImageNet \cite{deng2009imagenet} pretrained for the visual encoder. When our baseline is HTF \cite{WACV2023_Fedorishin}, we additionally consider the optical flow \cite{WACV2023_Fedorishin} for the attention module (more details are in Section 4.6). Our sound source localization framework is trained using the Adam optimizer \cite{kingma2014adam} with a learning rate of $10^{-4}$ and a batch size of 128. Following \cite{WACV2023_Fedorishin}, we train our model for 100 epochs for Flickr and VGG-Sound datasets. We use 4 synchronized RTX 3090 GPUs. The weights for $M_{final}$ in Eq. (5) are set as $w_1=w_2=w_3=1$. Also, we use $\lambda_1 = 1$, $\lambda_2 = 10$, $\tau = 0.03$, and $\delta = 25$ for our proposed loss functions ($\mathcal{L}_{avpm}$ and $\mathcal{L}_{sra}$).

	\begin{table}[t]
		\renewcommand{\tabcolsep}{0.2mm}
		\centering
		\caption{Experimental results on VGG-SS test set when the training sets are VGG-Sound10k and VGG-Sound144k, respectively.}
		\vspace{-0.3cm}
		\resizebox{0.9999\linewidth}{!}{
			\begin{tabular}{cccc}
				\Xhline{3\arrayrulewidth}
				\rule{0pt}{9.5pt} \bf Method & \bf Training Set & \bf cIoU$_{0.5}$$\uparrow$ & \bf AUC$\uparrow$ \\ \hline
				\rule{0pt}{10.pt}
				LVS \cite{CVPR2021_Chen} (CVPR'21)       & \multirow{4}{*}{\bf VGG-Sound10k} & 0.303 & 0.364  \\
				SSPL \cite{CVPR2022_Xuan} (CVPR'22)      &  & 0.314 & 0.369  \\
				Zhou et al. \cite{WACV2023_Zhou} (WACV'23)      &  & 0.350 & 0.376  \\
				HTF \cite{WACV2023_Fedorishin} (WACV'23)       &  & 0.393 & 0.398  \\\cdashline{1-4}
				\rule{0pt}{9.5pt}
				\bf Proposed Method        &  & \bf 0.403 & \bf 0.403 \\
				\Xhline{2\arrayrulewidth}
				\rule{0pt}{10.pt}
				LVS \cite{CVPR2021_Chen} (CVPR'21)       & \multirow{3}{*}{\bf VGG-Sound144k} & 0.344 & 0.382  \\
				SSPL \cite{CVPR2022_Xuan} (CVPR'22)      &  & 0.339 & 0.380  \\
				HTF \cite{WACV2023_Fedorishin} (WACV'23)       &  & 0.394 & 0.400  \\\cdashline{1-4}
				\rule{0pt}{9.5pt}
				\bf Proposed Method        &  & \bf 0.406 & \bf 0.405 \\
				\Xhline{3\arrayrulewidth}
		\end{tabular}}
		\label{table:param}
		\vspace{-0.3cm}
	\end{table}

	\subsection{Performance Comparison}
	
	We conduct the experiments to compare the effectiveness of our proposed method with the state-of-the-art sound source localization works \cite{CVPR2018_Senocak,CVPR2019_Hu_dmc,ECCV2020_Qian_coarsetofine,ECCV2020_afouras_avobject,CVPR2021_Chen,CVPR2022_Xuan,WACV2022_Shi,WACV2023_Zhou,WACV2023_Fedorishin}. Table 1 shows the performance of our method with the existing methods on Flickr-SoundNet. When the training set is Flickr10k, our method achieves 0.876 and 0.641 for cIoU$_{0.5}$ and AUC, respectively. Specifically, when compared to the HTF \cite{WACV2023_Fedorishin} which shows the highest performance among the existing methods, our method is 1.6\% higher for cIoU and 0.7\% higher for AUC metrices. Similar tendency is observed when the training set is Flickr144k training set. The experimental results on Table 1 demonstrate that our approach that considering the spatial knowledge of the audio-visual modalities and recursively refining the localization map leads to better localization performance.
	
	The experimental results on the VGG-Sound dataset are shown in Table 2. For the VGG-Sound Source test set, our method achieved improvements of 1.0\% cIoU and 0.5\% AUC in the VGG-Sound10k dataset, and 1.2\% cIoU and 0.5\% AUC in the VGG-Sound144k dataset over the HTF \cite{WACV2023_Fedorishin}. The results validate that our method outperforms existing methods and achieves a state-of-the-art performance over the existing sound source localization works.
	
	\subsection{Ablation Study}
	We conduct various ablation studies to investigate (1) effect of the proposed losses ({\it i.e.}, $\mathcal{L}_{avpm}$ and $\mathcal{L}_{sra}$), and (2) variation of the hyper-parameter $w_1$, $w_2$, $w_3$ for $M_{final}$. All the ablation studies are conducted using Flickr144k training set and Flickr test set. \\
	
	\noindent{\textbf{Effect of the Proposed Losses.}} We measure the performance by changing two types of the proposed losses $\mathcal{L}_{avpm}$ and $\mathcal{L}_{sra}$. The results are shown in Table 3. When each loss is considered, our method shows the improved performance agains the baseline in which those losses are not considered. When all the proposed losses are taken into account, we show the highest performance. By incorporating the proposed losses in the training phase, our method is able to learn more robust and discriminative features that are better suited for the sound source localization task. \\
	
	\noindent{\textbf{Variation of $\textit{\textbf{w}}_\textbf{1}$, $\textit{\textbf{w}}_\textbf{2}$, $\textit{\textbf{w}}_\textbf{3}$} We conduct additional ablation study to investigate the effect of our method to the parameters $w_1$, $w_2$, and $w_3$ as described in Section 3.3. The results of Table 4 show that the optimal results are obtained when $w_1$, $w_2$, and $w_3$ are set to 1. However, it's important to note that our method still outperforms existing methods even with different values for these parameters. These results suggest that the model is robust to parameter changes, but there may be an optimal combination that maximizes its effectiveness. In our future work, we are planning to build a framework that considers weight of the localization maps.
		
		\begin{table}[t]
			\centering
			\begin{center}
				\renewcommand{\tabcolsep}{2.9mm}
				\caption{Effect of the proposed audio-visual pair matching loss $\mathcal{L}_{avpm}$ and spatial region alignment loss $\mathcal{L}_{sra}$ on Flickr test set, where models are trained on the Flickr144k.}
				\vspace{-0.3cm}
				\resizebox{0.999\linewidth}{!}
				{
					\begin{tabular}{c cc cc}
						\Xhline{3\arrayrulewidth}
						\rule{0pt}{10pt} \textbf{Method} & $\mathcal{L}_{avpm}$  & $\mathcal{L}_{sra}$ & \bf cIoU$_{0.5}$$\uparrow$ & \bf AUC$\uparrow$ \\ \hline
						Baseline & -       & - & 0.865 & 0.642 \\\cdashline{1-5}
						\rule{0pt}{9.5pt} \multirow{3}{*}{\makecell{\bf Proposed\\ \bf Method}}  & \cmark  & - & 0.876 & 0.643 \\
						& -       & \cmark & 0.871 & 0.648 \\
						& \cmark       & \cmark & \textbf{0.881} & \textbf{0.652} \\\Xhline{3\arrayrulewidth}
				\end{tabular}}
				\label{table:ablation}
			\end{center}
			\vspace{-0.4cm}
		\end{table}
		
		\begin{table}[t]
			\centering
			\begin{center}
				\renewcommand{\tabcolsep}{2.5mm}
				\caption{Experimental results on Flickr test set according to the hyper-parameters $w_1$, $w_2$, and $w_3$ for $\textbf{M}_{final}$ in Eq. (5), where models are trained on the Flickr144k.}
				\vspace{-0.3cm}
				\resizebox{0.999\linewidth}{!}
				{
					\begin{tabular}{c cc cc}
						\Xhline{3\arrayrulewidth}
						$w_1$ \textbf{($\textbf{M}_v$)} & $w_2$ \textbf{($\textbf{M}_{a}$)} & $w_3$ \textbf{($\textbf{M}_v^{att}$)} & \bf cIoU$_{0.5}$$\uparrow$ & \bf AUC$\uparrow$ \\ \hline
						1 & 1 & 4 & 0.866 & 0.645 \\
						1 & 1 & 2 & 0.875 & 0.649 \\
						\textbf{1} & \textbf{1} & \textbf{1} & \textbf{0.881} & \textbf{0.652}\\
						1 & 2 & 1 & 0.871 & 0.639 \\
						2 & 1 & 1 & 0.876 & 0.650 \\
						2 & 2 & 1 & 0.871 & 0.643
						\\\Xhline{3\arrayrulewidth}
				\end{tabular}}
				\label{table:ablation}
			\end{center}
			\vspace{-0.2cm}
		\end{table}

		\begin{figure*}[t]
			\begin{minipage}[b]{1.0\linewidth}
				\centering
				\centerline{\includegraphics[width=17.8cm]{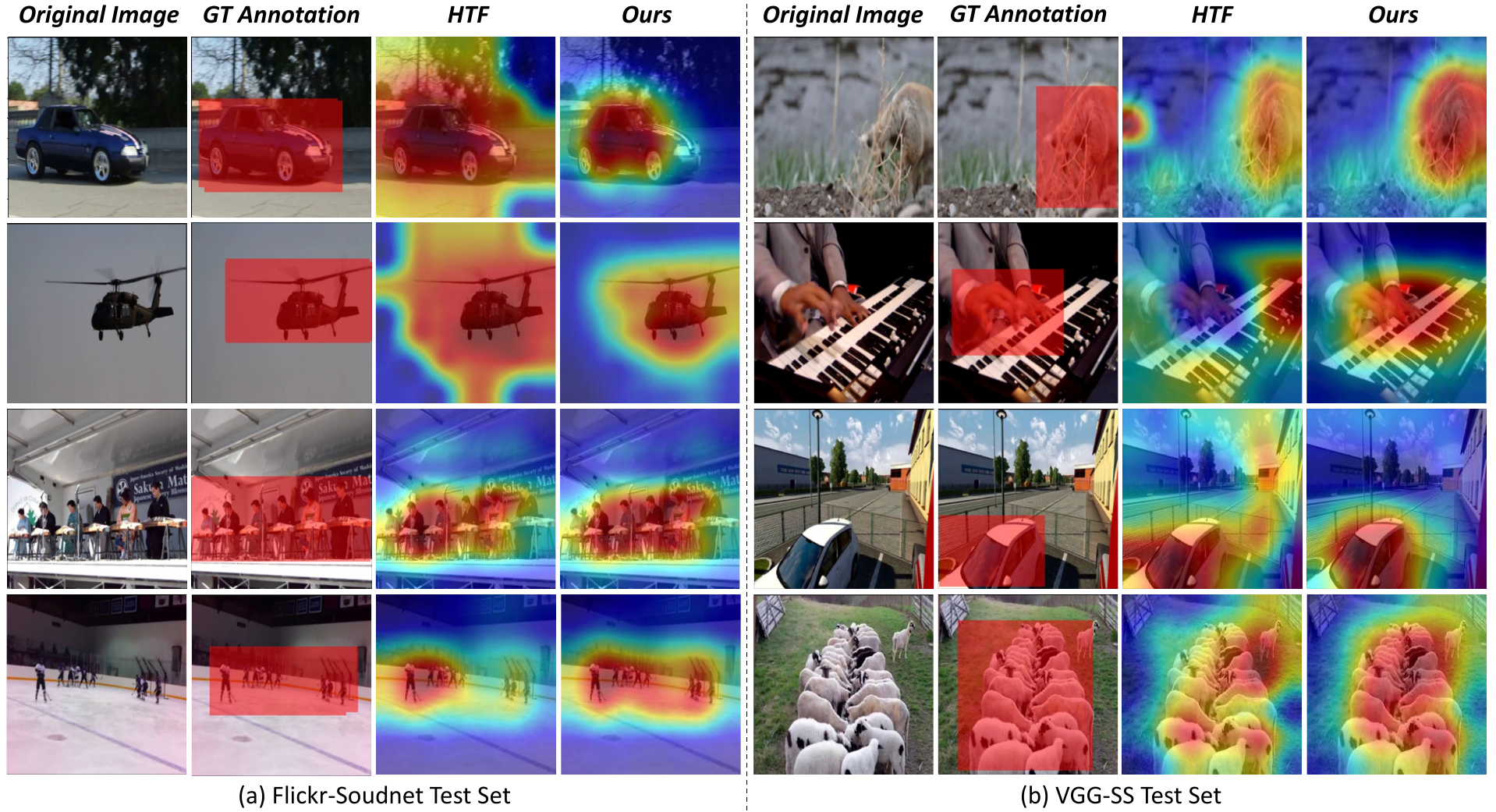}}
			\end{minipage}
			\vspace{-0.6cm}
			\caption{Visualization results for both (a) Flickr-SoundNet test set and (b) VGG-SS test set. Each result is obtained from models trained on the Flickr144k and VGG-Sound144k training sets. For the Flickr-SoundNet test set, annotation is created by combining bounding boxes from different annotators.}
			\vspace{-0.1cm}
		\end{figure*}

		\begin{figure*}[t]
			\begin{minipage}[b]{1.0\linewidth}
				\centering
				\centerline{\includegraphics[width=15.2cm]{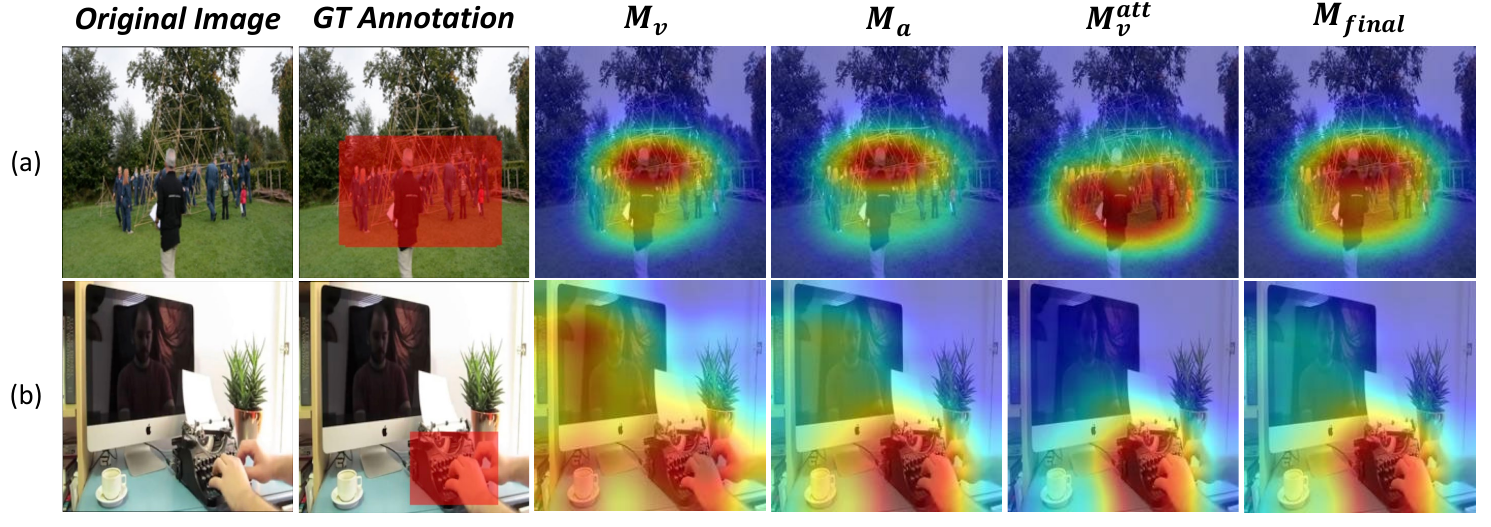}}
			\end{minipage}
			\vspace{-0.7cm}
			\caption{Visualization results of $\textbf{M}_v$, $\textbf{M}_a$, $\textbf{M}_v^{att}$, and the final localization map $\textbf{M}_{final}$ of our method on (a) Flickr-SoundNet test set and (b) VGG-SS test set.}
			\vspace{-0.2cm}
		\end{figure*}

		\begin{table}[t]
			\renewcommand{\tabcolsep}{0.1mm}
			\centering
			\caption{Experimental results on the cross-dataset evaluation. Note that we trained the model on the VGG-Sound10k and VGG-Sound144k and evaluated on the Flickr test set. `$\ast$' denotes our faithful reproduction of the method.}
			\vspace{-0.35cm}
			\resizebox{0.9999\linewidth}{!}{
				\begin{tabular}{cccc}
					\Xhline{3\arrayrulewidth}
					\rule{0pt}{9.5pt} \bf Method & \bf Training Set & \bf cIoU$_{0.5}$$\uparrow$ & \bf AUC$\uparrow$ \\ \hline
					\rule{0pt}{10.pt}
					LVS \cite{CVPR2021_Chen} (CVPR'21)       & \multirow{4}{*}{\bf VGG-Sound10k} & 0.618 & 0.536  \\
					SSPL \cite{CVPR2022_Xuan} (CVPR'22)      &  & 0.763 & 0.591  \\
					Zhou et al. \cite{WACV2023_Zhou} (WACV'23)       &  & 0.775 & 0.596  \\
					HTF \cite{WACV2023_Fedorishin}$^\ast$ (WACV'23)       &  &  0.842 & 0.628  \\\cdashline{1-4}
					\rule{0pt}{9.5pt}
					\bf Proposed Method        &  & \bf 0.875 & \bf 0.640 \\
					\Xhline{2\arrayrulewidth}
					\rule{0pt}{9.5pt}
					LVS \cite{CVPR2021_Chen} (CVPR'21)       & \multirow{3}{*}{\bf VGG-Sound144k} & 0.719 & 0.582  \\
					SSPL \cite{CVPR2022_Xuan} (CVPR'22)      &  & 0.767 & 0.605  \\
					HTF \cite{WACV2023_Fedorishin} (WACV'23)       &  & 0.848 & 0.640  \\\cdashline{1-4}
					\rule{0pt}{9.5pt}
					\bf Proposed Method       &  & \bf 0.881 & \bf 0.651 \\
					\Xhline{3\arrayrulewidth}
				\end{tabular}
			}
			\label{table:param}
			\vspace{-0.05cm}
		\end{table}

		\begin{table}[t]
			\renewcommand{\tabcolsep}{0.9mm}
			\centering
			\caption{Experimental results on Flickr test set with respect to various sound source localization frameworks.}
			\vspace{-0.35cm}
			\resizebox{0.9999\linewidth}{!}{
				\begin{tabular}{cccc}
					\Xhline{3\arrayrulewidth}
					\rule{0pt}{9.5pt} \bf Method & \bf Training Set & \bf cIoU$_{0.5}$$\uparrow$ & \bf AUC$\uparrow$ \\ \hline
					\rule{0pt}{10.pt}
					LVS \cite{CVPR2021_Chen} (CVPR'21)       & \multirow{2}{*}{\bf Flickr144k} & 0.699 & 0.573  \\
					\textbf{Proposed Method (LVS)}  & & \bf 0.718 & \bf 0.577 \\\hline
					\rule{0pt}{9.5pt}
					HTF \cite{WACV2023_Fedorishin} (WACV'23)      & \multirow{2}{*}{\bf Flickr144k} & 0.865 & 0.639  \\
					\textbf{Proposed Method (HTF)} & & \bf 0.881 & \bf 0.652 \\
					\Xhline{3\arrayrulewidth}
				\end{tabular}
			}
			\label{table:param}
			\vspace{-0.05cm}
		\end{table}
		
		\begin{table}[t]
			\renewcommand{\tabcolsep}{1.3mm}
			\centering
			\caption{Experimental results on Flickr test set using mcIoU metric.}
			\vspace{-0.35cm}
			\resizebox{0.9999\linewidth}{!}{
				\begin{tabular}{cccc}
					\Xhline{3\arrayrulewidth}
					\rule{0pt}{9.5pt} \bf Method & \bf Training Set & \bf cIoU$_{0.5}$$\uparrow$ & \bf mcIoU$\uparrow$ \\ \hline
					\rule{0pt}{10.pt}
					LVS \cite{CVPR2021_Chen} (CVPR'21)       & \multirow{2}{*}{\bf Flickr144k} & 0.699 & 0.231  \\
					HTF \cite{WACV2023_Fedorishin} (WACV'23)       &  & 0.865 & 0.363  \\\cdashline{1-4}
					\rule{0pt}{9.5pt}
					\bf Proposed Method        &  & \bf 0.881 & \bf 0.381 \\
					\Xhline{3\arrayrulewidth}
				\end{tabular}
			}
			\label{table:param}
			\vspace{-0.2cm}
		\end{table}
		
		\subsection{Visualization Results}
		
		We compare our method with the current state-of-the-art approach, HTF \cite{WACV2023_Fedorishin}, by visualizing their sound source localization results on the Flickr-SoundNet and VGG-SS test set. The results are shown in Figure 3. Through the visualization results, our method can accurately localize the sound-making objects (GT annotation indicates the region of the sound-making objects). Since our method considers the spatial information of both audio and visual modalities and recursively updates the localization map, more precise attention maps are obtained.
		
		Furthermore, Figure 4 shows visualization results of the various localization maps $\textbf{M}_v$, $\textbf{M}_a$, $\textbf{M}_v^{att}$, and $\textbf{M}_{final}$ of our method. The visualization results show that considering audio-visual localization map and recursively updating them contributes to $M_{final}$ for concentrating on a more accurate location. By doing so, our method shows the improving performance.

		\subsection{Discussions}
		
		\noindent{\textbf{Cross Dataset Evaluation.}}  
		To show the effectiveness of our method on the cross dataset environment, we train our model on the VGG-Sound training set and evaluate it on the Flickr-SoundNet test set. This cross-dataset evaluation enables us to assess the ability of model to generalize and to check to new and diverse data sources. The results of Table 5 show the results when the training sets are VGG-Sound10k and VGG-Sound144k, respectively. The results show that the performances of our method are significantly higher than the existing methods. As a result, our model demonstrates the potential to demonstrate sufficient generalization capabilities essential for real-world applications involving diverse data sources.\\

		\noindent{\textbf{Generalization Ability of Our Method.}} In this subsection, we conduct experiments to see the generalization ability of our method by varying the baseline. To this end, we adopt the two baselines: LVS \cite{CVPR2021_Chen} and HTF \cite{WACV2023_Fedorishin}. The results are shown in Table 6. All the methods are trained with Flickr144k and evaluated on Flickr test set. 
		As shown in the table, when our baseline is LVS \cite{CVPR2021_Chen}, we achieve 1.19\% cIoU and 0.4\% AUC improvement compared to the original LVS. The results on HTF \cite{WACV2023_Fedorishin} also show a similar tendency. The results indicate that our method has broader applicability and can be integrated with various sound source localization frameworks. \\

		\noindent{\textbf{Evaluation on the Proposed mcIoU Metric.}} Note that consensus intersection over union (cIoU) \cite{CVPR2018_Senocak} metric has been widely used for comparing sound source localization methods. In this subsection, we additionally introduce a new metric called mcIoU (mean cIoU) to investigate the performance while varying the IoU threshold. For calculating mcIoU metric, we take the average cIoU across IoU threshold 0.5:0.05:0.95. The results are shown in Table 7. Compared to the existing methods \cite{CVPR2021_Chen,WACV2023_Fedorishin}, the performances of our method consistently improved. The results demonstrate that the effectiveness of our method considers spatial cues of audio modality and performs sound source localization in a recursive manner.\\
		
		\noindent{\textbf{Computational Costs.}} In this section, we compare training time, inference time, and the number of parameters. It is shown in Table 8. We compare our method with HTF \cite{WACV2023_Fedorishin} which shows the highest performance among the existing methods. Since our method adopts the recursive method, the training time, inference time, and the number of parameters of our method are slightly increased (3.38\%, 7.69\% and 1.92\% for training, inference, and parameters, respectively). Nevertheless, we claim that the increased times of training and inference time and the number of parameters are marginal compared to the HTF \cite{WACV2023_Fedorishin}.

		\begin{table}[t]
			\renewcommand{\tabcolsep}{0.7mm}
			\centering
			\caption{The comparisons of training time, inference time, and the number of parameters.}
			\vspace{-0.2cm}
			\resizebox{0.9999\linewidth}{!}{
				\begin{tabular}{cccc}
					\Xhline{3\arrayrulewidth}
					\rule{0pt}{9.5pt} \bf \multirow{2}{*}{\bf Method} & \bf Training (s) & \bf Inference (s) & \multirow{2}{*}{\bf \#params}\\
					& \bf (\textit{per iter}) & \bf (\textit{per image})
					\\ \hline
					\rule{0pt}{10.pt}
					HTF \cite{WACV2023_Fedorishin} (WACV'23)      & 0.385 & 0.039 & 33.85M  \\
					\textbf{Proposed Method} & 0.398 & 0.042 & 34.50M \\
					\Xhline{3\arrayrulewidth}
				\end{tabular}
			}
			\label{table:param}
			\vspace{-0.4cm}
		\end{table}
		
		\section{Conclusion}
		
		In this paper, we propose a novel sound source localization framework that considers the inherent spatial information of the audio modality as well as the visual modality for exploiting more abundant spatial knowledge. To this end, our framework consists of two stages: (1) audio-visual spatial integration network and (2) recursive attention network. The audio-visual spatial integration network is designed to incorporate the spatial information of the audio-visual modalities. By focusing on the attention region generated by the audio-visual spatial integration network, the recursive attention network aims to perform more precise sound source localization. At this time, we devise audio-visual pair matching loss and spatial region alignment loss to effectively guide the features from the audio-visual modalities to resemble the features of the attentive information. We believe that our approach, integrating spatial knowledge of audio-visual modalities and recursively refining the results leads to more improved accuracy and it can be utilized in various practical applications.
		
		\begin{acks}
			This work was supported in part by the National Research Foundation of Korea (NRF) grant funded by the Korea government (MSIT) (No. RS-2023-00252391) and by Institute of Information \& communications Technology Planning and Evaluation (IITP) grant funded by the Korea government (MSIT) (No. 2022-0-00124: Development of Artificial Intelligence Technology for Self-Improving Competency-Aware Learning Capabilities, No. RS-2022-00155911: Artificial Intelligence Convergence Innovation Human Resources Development (Kyung Hee University)) and by the MSIT (Ministry of Science and ICT), Korea, under the National Program for Excellence in SW (2023-0-00042) supervised by the IITP in 2023.
		\end{acks}

		\bibliographystyle{ACM-Reference-Format}
		\bibliography{ACMMM_references_camready}


\clearpage

\setcounter{section}{0}
\setcounter{figure}{0}
\setcounter{table}{0}
\setcounter{equation}{0}

\appendix

\noindent\textbf{\huge Supplementary Material}
\vspace{3mm}





This manuscript provides the additional results of the proposed method. Section A shows our additional implementation details, Section B indicates the additional experimental results to show the effectiveness of the cross dataset evaluation and proposed two modules (spatial integration and recursive attention), and Section C shows the additional visualization results, respectively. Please note that [PXX] indicates the reference in the main paper.

\section{Additional Implementation Details}

As mentioned in the main paper, we adopt the ResNet-18[P20] as the audio encoder. However, to match the input size, we adjusted the input channel of the first convolutional network to 1 and the output channel to 64, employing a kernel size of 7, stride 2, and padding 3. In addition, we used the Adam optimizer for training, with parameters $(\beta_1, \beta_2) = (0.9, 0.999)$, and a learning rate of 0.001. These are standard values for Adam and provided stable training dynamics in our experiments.

\section{Additional Experiments}

\noindent {\textbf{Cross Dataset Evaluation (Train: Flickr, Test: VGG-SS).}}
To further validate the generalization ability of our method, we conducted an experiment where we trained our model on the Flickr dataset [P4] and evaluated it on the VGG-SS dataset [P9]. This is in contrast to our main paper, where we trained on the VGG-SS dataset and evaluated it on the Flickr dataset. As depicted in Table 1, the results demonstrate that our model still outperforms the HTF [P15], which shows state-of-the-art performance. These results confirm the ability of our method to generalize to new and diverse sources, further supporting the robustness of our approach.\\

\noindent {\textbf{Performance of proposed two modules (spatial integration and recursive attention).}}
We conduct the additional ablation study on Flicker144k [P4] and VGG-Sound144k [P9] datasets when the proposed two modules (spatial integration and recursive attention) are considered or not. We compare the performances with two state-of-the-art methods, SSPL [P60] and HTF [P15]. The results are shown in Table 2. When the spatial integration network and the recursive attention network are considered individually, our method already exhibits improved performances compared to the existing methods. It demonstrates that each module contributes to the sound source localization task. Moreover, when the two modules are considered together, our method obtains further performance improvement.

\begin{table}[t]
    \renewcommand{\tabcolsep}{1.2mm} 
    \centering
    \caption{Experimental results on the cross dataset evaluation. Note that we trained the model on the Flickr10k [P4] and Flickr144k [P4] and evaluated on the VGG-SS test set [P9]. `$\ast$' denotes our faithful reproduction of the method. }
        \vspace{-0.2cm}
	\resizebox{0.9999\linewidth}{!}{
		\begin{tabular}{cccc}
		\Xhline{3\arrayrulewidth}
		\rule{0pt}{9.5pt} \bf Method & \bf Training Set & \bf cIoU$_{0.5}$$\uparrow$ & \bf AUC$\uparrow$ \\ \hline
		\rule{0pt}{10.pt}
			HTF [P15]$^\ast$ (WACV'23)       & \bf Flickr10k &  0.384 & 0.396  \\\cdashline{1-4}
                \rule{0pt}{9.5pt}
			\bf Proposed Method        &  & \bf 0.396 & \bf 0.400 \\
        \Xhline{2\arrayrulewidth}
        \rule{0pt}{9.5pt}
			HTF [P15]$^\ast$ (WACV'23)       &  \bf Flickr144k & 0.385 & 0.396  \\\cdashline{1-4}
                \rule{0pt}{9.5pt}
			\bf Proposed Method       &  & \bf 0.399 & \bf 0.401 \\
        \Xhline{3\arrayrulewidth}
	\end{tabular}
}
    \label{table:param}
    \vspace{-0.3cm}
\end{table}

\begin{table}[t]
    \renewcommand{\tabcolsep}{1.2mm}
    \centering
    \caption{Experiments result for performance comparison of two modules. Note that `SI' represents Spatial Integration, and `RA' represents Recursive Attention.}
        \vspace{-0.3cm}
	\resizebox{0.9999\linewidth}{!}{
		\begin{tabular}{cccc}
		\Xhline{3\arrayrulewidth}
		\rule{0pt}{9.5pt} \bf Method & \bf Training Set & \bf cIoU$_{0.5}$$\uparrow$ & \bf AUC$\uparrow$\\ \hline
		\rule{0pt}{8.pt}
            SSPL [P60] (CVPR'22)       & \multirow{5}{*}{\bf Flickr144k} & 0.759 & 0.610  \\
            HTF [P15] (WACV'23)     &  & 0.865 & 0.639  \\
            \bf Ours (w/ SI, w/o RA)	      &  & \bf 0.870 & \bf 0.645  \\
			\bf Ours (w/o SI, w/ RA)       &  & \bf 0.870 & \bf 0.649  \\
            \bf Ours (w/ SI, w/ RA)       &  & \bf 0.881 & \bf 0.652 \\
        \Xhline{2\arrayrulewidth}
        \rule{0pt}{8. pt}
            SSPL [P60] (CVPR'22)       & \multirow{5}{*}{\bf VGG-Sound144k} & 0.339 & 0.380  \\
            HTF [P15] (WACV'23)     &  & 0.394 & 0.400 \\
            \bf Ours (w/ SI, w/o RA)	      &  & \bf 0.403 & \bf 0.402  \\
			\bf Ours (w/o SI, w/ RA)       &  & \bf 0.404 & \bf 0.403  \\
            \bf Ours (w/ SI, w/ RA)       &  & \bf 0.406 & \bf 0.405  \\
        \Xhline{3\arrayrulewidth}
	\end{tabular}}
    \label{table:param}
    \vspace{-0.3cm}
\end{table}


\begin{figure*}[t]
    \begin{minipage}[b]{1.0\linewidth}
	\centering
	\centerline{\includegraphics[width=18.0cm]{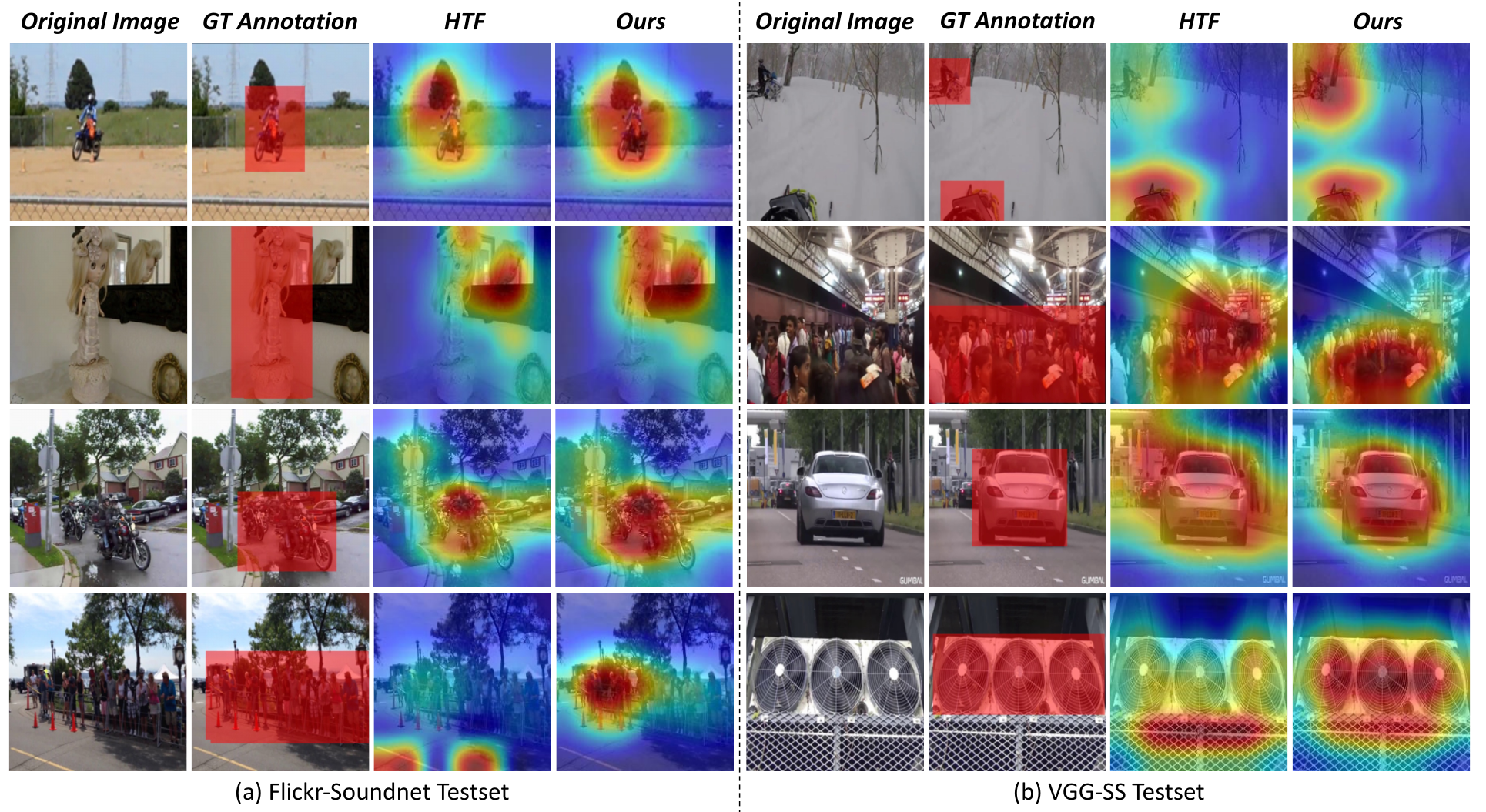}}
	\end{minipage}
    \vspace{-0.6cm}
    \caption{Expanded visualization results for both (a) Flickr-SoundNet test set and (b) VGG-SS test set. Each result is obtained from models trained on the Flickr144k and VGG-Sound144k training sets.}
    \vspace{-0.1cm}
\end{figure*}

\begin{figure*}[t]
    \begin{minipage}[b]{1.0\linewidth}
	\centering
	\centerline{\includegraphics[width=16cm]{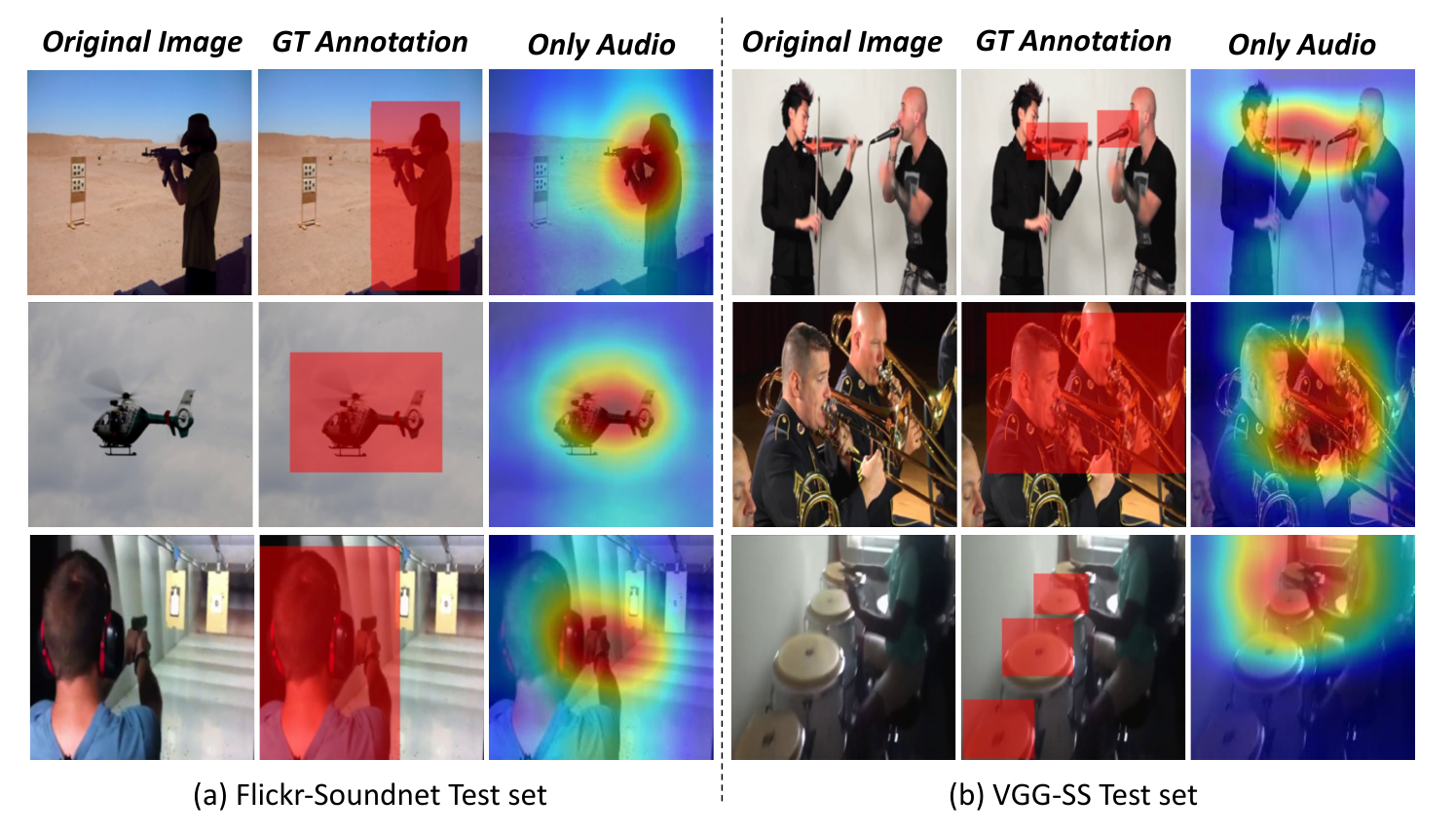}}
	\end{minipage}
    \vspace{-0.6cm}
    \caption{Visualization results for both (a) Flickr-SoundNet test set and (b) VGG-SS test set with only audio. Each result is obtained from models trained on the Flickr144k and VGG-Sound144k training sets.}
    \vspace{-0.1cm}
\end{figure*}

\section{Additional Visualizations}

\noindent {\textbf{Sound Source Localization Result Comparisons of HTF [P15] and Ours.}} We additionally present a comparison between our method and HTF [P15] in sound source localization results. We used the codes provided by the authors to obtain the HTF results, which are presented in Figure 1. The fourth row of Figure 1(a) and Figure 1(b) show that the HTF model failed to localize the sound-making regions while our method was able to focus on them. The results demonstrate that the proposed method, which considers the spatial information of the audio-visual modalities and improves the localization in a recursive manner, outperforms the HTF. \\

\noindent {\textbf{Spatial Information of the Audio Modality.}} Furthermore, we provide the sound source localization results with only audio features to investigate the effect of the spatial information of the audio modality. The outcomes are shown in Figure 2, revealing that the audio modal itself also contains valuable spatial cues for inferring the sound-making objects. \\

\noindent {\textbf{Video Results of the Proposed Sound Source Localization.}} In addition to the visualizations provided in the paper, we include supplementary video material to show the results of our method. The video displays the outcomes of our method applied to some samples from both the Flickr-Soundnet and VGG-SS datasets. Please see \href{https://github.com/VisualAIKHU/SIRA-SSL}{\color{red}{https://github.com/VisualAIKHU/SIRA-SSL}}.

\clearpage
\end{document}